\documentclass{iopart}

	\usepackage{iopams}
	\usepackage{setstack}
	\usepackage{graphicx}
	\usepackage{subfigure}

\begin{document}

\title[Thermodynamical instabilities in General Relativity]{Thermodynamical instabilities of perfect fluid spheres in General Relativity}
\author{Zacharias Roupas$^{1,2}$}
\address{$^1$ Institute of Nuclear and Particle Physics, N.C.S.R. Demokritos, GR-15310 Athens, Greece} 
\address{$^2$ Physics Department, National Technical University of Athens, GR-15780, Athens, Greece}
\ead{roupas@inp.demokritos.gr}

\begin{abstract}
	For a static, perfect fluid sphere with a general equation of state, we obtain the relativistic equation of hydrostatic equilibrium, namely the Tolman-Oppenheimer-Volkov equation, as the thermodynamical equilibrium in the microcanonical, as well as the canonical, ensemble. We find that the stability condition determined by the second variation of entropy coincides with the dynamical stability condition derived by variations to first order in the dynamical Einstein's equations. Thus, we show the equivalence of microcanonical thermodynamical stability with linear dynamical stability for a static, spherically symmetric field in General Relativity. We calculate the Newtonian limit and find the interesting property, that the microcanonical ensemble in General Relativity transforms to the canonical ensemble for non-relativistic dust particles. Finally, for specific kinds of systems, we study the effect of the cosmological constant to the microcanonical thermodynamical stability of fluid spheres.
\end{abstract}
%\pacs{}

\maketitle

\section{Introduction}

\indent The deep connection between gravity and thermodynamics is being extensively studied for several decades now, starting from the four laws of black hole mechanics \cite{bekenstein,bch} and Hawking radiation \cite{hawking}. These laws were derived by Einstein's equations. More recently, by reversing the logic around, many attempts have been made to derive Einstein's equations from thermodynamics \cite{jacobson,padmanabhan,verlinde} and thus realize gravity as an emergent phenomenon, rooted in a deeper, underlying and yet unknown, microscopic structure. All these attempts were based on thermodynamics on an horizon or on some other boundary, making always use of some form of holographic principle, pointing in this way to a quantum gravity theory. However, we think it is crucial, before introducing some sort of exotic or quantum principle, to understand at purely classical level \cite{hu}, as deeply as possible, the thermodynamics of self-gravitating gas. In this way, it will be possible to figure out which peculiar properties of gravitational thermodynamics are of quantum nature, rooted in a deeper theory, and which are just built-in features of classical General Relativity. In this spirit, we will study the classical thermodynamics of self-gravitating gas in General Relativity with no reference to the thermodynamics on an horizon or a screen. \\
\indent In fact, the attempt to derive Einstein's equations from thermodynamics of self-gravitating gas dates back to $1965$ in a work of Cocke \cite{cocke}. He proposed a maximum entropy principle for a static, spherically symmetric, perfect fluid to derive the relativistic equation of hydrostatic equilibrium, that is widely known as Tolman-Oppenheimer-Volkof (TOV) equation. In his formulation the fluid was in adiabatic motion so that the total entropy was constant and the initial data constraint equation together with the radial-radial component of Einstein's equations were used. Later, in 1981, Sorkin, Wald and Zhang (SWZ) \cite{swz} developed a different entropy principle for radiation, which did not need the adiabatic condition and used only the constraint equation. Very recently, Gao \cite{gao} generalized SWZ's formulation for an arbitrary perfect fluid. However, in Gao's \cite{gao} formulation is not clear in which thermodynamic ensemble the calculation is performed. It is crucial to specify the ensemble, because in Gravity the thermodynamic ensembles are \textit{not equivalent} \cite{padman,katz} regarding stability properties. We will sharpen his argument, by deriving the TOV equation from thermodynamics, working clearly both in the microcanonical as well as the canonical ensemble and in a hopefully, more straightforward and simply way. \\
\indent Most importantly, we show that gravitation is equivalent to thermodynamics in the microcanonical ensemble, not only regarding the equilibria (i.e. TOV  equation), but even at the level of \textit{stability}. Indeed, we find that linear dynamical stability, defined by Einstein's equations, coincides with microcanonical, thermodynamical stability, defined by the second variation of entropy, for a static, perfect fluid sphere with a general equation of state. In addition, we show that the microcanonical ensemble in General Relativity, becomes the canonical ensemble in the limit of non-relativistic, dust matter. \\
\indent Finally, we study the effect of the cosmological constant on the stability of fluid spheres. This study can be considered as the relativistic generalization of previous studies of self-gravitating gas in the Newtonian limit \cite{agr,agrN,agrP}. We perform the analysis for matter and radiation with a linear equation of state with the assumptions of constant entropy per baryon and chemical composition throughout the whole fluid. A general result is that an increase in the cosmological constant tends to stabilize the system. In addition, the critical minimum radius, down to which there exist equilibrium configurations, `hits' the cosmological horizon at some value of the cosmological constant. \\
\indent Our work is organized as follows. In section \ref{sec:dynamical} we derive the condition for linear dynamical stability from the Einstein's equations. In section \ref{sec:first} we prove that TOV equation can be derived from the microcanonical ensemble, i.e. by the first variation of entropy with fixed mass and number of particles, and we discuss the canonical ensemble as well. In section \ref{sec:thermo} we calculate the second variation of entropy and prove that the condition for microcanonical, thermodynamical stability is equivalent to the condition for linear dynamical stability calculated in section \ref{sec:dynamical}. In section \ref{sec:newtonian} we calculate the Newtonian limit of the condition for microcanonical thermodynamical stability, found in section \ref{sec:thermo}. Finally, in section \ref{sec:lambda} we study the effect of the cosmological constant to the stability of fluid spheres with specific equations of state.

\section{Linear dynamical stability}\label{sec:dynamical}

\indent In case of a perfect fluid that is static and spherically symmetric, the Einstein equations reduce to the equation known as Tolman-Oppenheimer-Volkoff (TOV) \cite{tolman,ov} (see \cite{boehmer} for a study on TOV with a cosmological constant). TOV is the equation that describes hydrostatic equilibria. In \ref{app:A} we review how this equation is derived from Einstein's equations. In this section we want to find the condition for linear dynamical stability, i.e. for stability under radial variations, to first order, about the equilibrium. This problem has been studied by Yabushita \cite{yabushita}. In \ref{app:A} we review analytically the calculations of Yabushita, including a cosmological constant, as well. Our final purpose is to compare dynamical stability with thermodynamical stability, which we will finally do in section \ref{sec:thermo}. For this we need the expression for linear dynamical stability of a general equation of state which we calculate in the followings. \\
\indent The metric for a spherically symmetric system can be written in the form \cite{tolman}
\begin{equation}\label{eq:metric}
	ds^2 = e^\nu c^2dt^2 - e^\lambda dr^2 - r^2 (d\theta^2 + sin^2\theta d^2\phi )
\end{equation}
with $\nu = \nu (r,t)$ and $\lambda = \lambda (r,t)$. For a perfect fluid, the energy momentum tensor is
\begin{equation}\label{eq:enmom}
	T^\mu_\nu = (p + \rho c^2)g_{\alpha \nu} \frac{dx^\alpha}{ds}\frac{dx^\nu}{ds} - p \delta^\mu_\nu
\end{equation}
which at the equilibrium becomes just
\begin{equation}
	T^\mu_\nu = \mbox{diag}(\rho_e c^2,-p_e,-p_e,-p_e)
\end{equation}
where the suffix $e$ denotes quantities at the equilibrium. At the equilibrium the metric is static and the Einstein's equations reduce to just two equations (see \ref{app:A}):
\begin{eqnarray}
\fl	\label{eq:TOV}
{p_e}' = -(\frac{p_e}{c^2} + \rho_e)\left( \frac{Gm_e(r)}{r^2} + 4\pi G \frac{p_e}{c^2}r - \frac{8\pi G}{3} \rho_\Lambda r\right)
	\left( 1 - \frac{2Gm_e(r)}{rc^2} - \frac{8\pi G}{3c^2} \rho_\Lambda r^2\right)^{-1}		\\
\label{eq:massd}
m_e' = 4\pi r^2 \rho_e
\end{eqnarray}
where $\rho_\Lambda$ is the density corresponding to a (positive or negative) cosmological constant $\Lambda$, defined as:
\begin{equation}\label{eq:lambda}
	\rho_\Lambda = \frac{\Lambda c^2}{8\pi G}
\end{equation}
Equation (\ref{eq:TOV}) is the TOV equation that defines hydrostatic equilibria, while equation (\ref{eq:massd}) implies that $m(r)$ is the total mass-energy inside radius $r$ \cite{weinberg}. The equilibria can be determined by these two equations, provided an equation of state $p = p(\rho)$ and initial conditions $p(0)$, $m(0)$ are given. \\
\indent Now, let us apply small perturbations about the equilibria (see \ref{app:A}) as in equation \eref{eq:pert}. Let 
\begin{equation}\label{eq:f}
	f(r,t) \equiv \delta m(r,t)
\end{equation}
with
\begin{equation}\label{eq:drho}
	\delta \rho = \frac{1}{4\pi r^2}\frac{\partial f}{\partial r}
\end{equation}
Then, assuming a perturbation
\begin{equation}\label{eq:dm}
	\delta m \sim e^{\sigma t}
\end{equation}
with $\delta m(r=0) = 0$ one obtains from Einstein's equations, to first order, that:
\begin{equation}\label{eq:yabushita}
 \delta p' + \frac{{\nu_e}'}{2}(\delta p + \delta \rho c^2)
		+\frac{{\delta\nu}'}{2}(p_e + \rho_e c^2)  = \frac{e^{\lambda_e - \nu_e}}{4\pi r^2}\sigma^2 f
\end{equation}
This is equation $(2.13)$ in Yabushita \cite{yabushita} and is proved in \ref{app:A} with a cosmological constant included in the calculations. It is clear from equation \eref{eq:dm} that if $\sigma^2 < 0$ the system will pulsate about the equilibrium and therefore it is stable. Given an equation of state, all variations in equation \eref{eq:yabushita} can be expressed with respect to $f$, so that \eref{eq:yabushita} becomes a differential equation with respect to $f$ and a differential operator may be defined on the left hand side as we will see. The stability is therefore determined by the sign of the eigenvalues of this operator. If it has any positive eigenvalue the system is unstable. \\
\indent To first order in perturbations \eref{eq:pert}, equation \eref{eq:tensorcom2} becomes
\begin{equation}\label{eq:temp1}
	\frac{8\pi G}{c^4}\delta p = \left(1 - \frac{2Gm}{rc^2} - \frac{8\pi G}{3c^2}\rho_\Lambda r^2\right)
		\left\lbrace \frac{\delta\nu '}{r} - \delta\lambda\left( \frac{\nu'}{r} + \frac{1}{r^2}\right)\right\rbrace
\end{equation}
The perturbation $\delta \lambda$ can be calculated by equation (\ref{eq:lambda_m}) to be: 
\[
	\delta \lambda = \left(1 - \frac{2Gm}{rc^2} - \frac{8\pi G}{3c^2}\rho_\Lambda r^2\right)^{-1}\frac{2G}{rc^2}f
\]
Solving equation (\ref{eq:temp1}) with respect to $\delta \nu'$ and substituting it together with $\nu'$, from equation \eref{eq:tolequi1}, in equation \eref{eq:yabushita} we get:
\begin{eqnarray}\label{eq:instgen}
\fl	 \delta p' -\frac{{p_e}'}{p_e + \rho_e c^2}(\delta p + \delta\rho c^2) + 
  		(p_e + \rho_e c^2)\left(1 - \frac{2Gm}{rc^2} - \frac{8\pi G}{3c^2}\rho_\Lambda r^2\right)^{-1}  \nonumber \\
  		\times \left\lbrace \frac{4\pi G}{c^4} \delta p  + \frac{G}{rc^2}\left(-2\frac{{p_e}'}{p_e+\rho_e c^2} + \frac{1}{r} \right)f \right\rbrace = \frac{e^{\lambda_e - \nu_e}}{4\pi r^2}\sigma^2 f
\end{eqnarray}
This is an important equation for our purposes, since it depends only on $\delta p$ and $f=\delta m$. This enables us to compare it with the corresponding expression deduced from thermodynamical stability and in this way determine the necessary condition for the equation of state, so that the two kinds of stability to be equivalent. This will be done in section \ref{sec:thermo}. \\
\indent Let us assume a general equation of state of the type
\[
	p = p(\rho)
\]
Let 
\begin{equation}
	g(r) = \frac{1}{c^2}\frac{\partial p}{\partial\rho}
\end{equation}
so that 
\begin{equation}\label{eq:ppr}
	p' = g(r)c^2\rho'\quad\mbox{and}\quad \delta p = g(r)c^2\delta\rho = \frac{gc^2}{4\pi r^2}\frac{df}{dr}
\end{equation}
and
\begin{equation}\label{eq:dppr}
	{\delta p}' = \frac{c^2}{4\pi r^2}\left(g\frac{d^2f}{dr^2} + \left(g' - \frac{2g}{r}\right)\frac{df}{dr}\right)
\end{equation}
Substituting equations \eref{eq:ppr}, \eref{eq:dppr} and \eref{eq:drho} into \eref{eq:instgen} we finally get:
\begin{eqnarray}\label{eq:dyneig}
	gc^2\left\lbrace\frac{d^2f}{dr^2} + \left[ \frac{g'}{g} - \frac{2}{r} - \frac{(g+1)c^2{\rho_e}'}{p_e+\rho_e c^2} 
		+ \frac{4\pi G}{c^4}r \right.\right.
		\nonumber \\
\fl		\left.\left.
	\times\left(1 - \frac{2Gm}{rc^2} - \frac{8\pi G}{3c^2}\rho_\Lambda r^2\right)^{-1} (p_e+\rho_e c^2)\right]
			\frac{df}{dr}\right\rbrace
			+\frac{4\pi G}{c^2} r(p_e+\rho_e c^2)  \nonumber \\
\fl			\times\left(1 - \frac{2Gm}{rc^2} - \frac{8\pi G}{3c^2}\rho_\Lambda r^2\right)^{-1} \left( -\frac{2gc^2{\rho_e}'}{p_e+\rho_e c^2} + \frac{1}{r}\right)f
 = e^{\lambda_e - \nu_e}\sigma^2 f
\end{eqnarray}
Thus, linear dynamical stability is determined by the sign of the eigenvalues of the operator
\begin{eqnarray}\label{eq:dynoper} 
\fl	\hat{L} = gc^2\left\lbrace\frac{d^2}{dr^2} + \left[ \frac{g'}{g} - \frac{2}{r} - \frac{(g+1)c^2{\rho_e}'}{p_e+\rho_e c^2} 
		+ \frac{4\pi G}{c^4}r 
		\right.\right.\nonumber\\
		\times \left.\left.\left(1 - \frac{2Gm}{rc^2} - \frac{8\pi G}{3c^2}\rho_\Lambda r^2\right)^{-1}(p_e+\rho_e c^2)\right]\frac{d}{dr} \right\rbrace \nonumber \\
\fl  + \frac{4\pi G}{c^2} 
 (p_e+\rho_e c^2) r
	\left(1 - \frac{2Gm}{rc^2} - \frac{8\pi G}{3c^2}\rho_\Lambda r^2\right)^{-1} \left( -\frac{2gc^2{\rho_e}'}{p_e+\rho_e c^2} + \frac{1}{r}\right) 
\end{eqnarray}
with $g = \partial p/\partial\rho c^2$. If there exists any positive eigenvalue the equilibrium is unstable. In a series of equilibria, a turning point of stability happens at the equilibrium for which:
\begin{equation}
	\hat{L}f = 0
\end{equation}

\section{General relativistic equation of hydrostatic equilibrium from thermodynamics}\label{sec:first}

\indent As we have stated in the introduction, Gao \cite{gao} generalized SWZ's \cite{swz} formulation and managed to derive TOV equation by extremizing the entropy, for an arbitrary perfect fluid, but with no reference to any specific thermodynamic ensemble. We know that specifying the ensemble is very important in gravity, because they are \textit{not} equivalent. In this section we will derive the TOV equation from thermodynamics, working clearly in the microcanonical ensemble and performing the calculation in a different, straightforward way. We will include in the calculation a cosmological constant, as well. Next, we will discuss the canonical ensemble, too. \\
\indent Following Gao \cite{gao} and SWZ \cite{swz}, let us state the main assumptions: 
\begin{itemize}
\item We assume a perfect fluid for which the first law of thermodynamics holds:
\begin{equation}\label{eq:1stlaw}
	ds = \frac{c^2}{T}d\rho - \frac{\mu}{T}dn
\end{equation}
where $s$ is the entropy per unit volume, $T$ is the temperature, $\rho$ is the mass density, related with the energy density by $\epsilon = \rho c^2$, $\mu$ is the chemical potential and $n$ is the baryon number density. This equation implies that $\rho$ and $n$ can be considered as two independent variables and $s = s(\rho,n)$.
\item The Gibbs-Duhem relation, which is a general expression for a perfect fluid, holds:
\begin{equation}\label{eq:gibbs}
	T s = \rho c^2 + p - \mu n
\end{equation}
\item The system is spherically symmetric and the entropy maxima necessarily correspond to static configurations\footnote{Note that this is proved by SWZ \cite{swz} for the asymptotically flat case.}. Therefore, the initial value constraint equation \cite{swz} for time symmetric data holds:
\begin{equation}\label{eq:ivalue}
	^{(3)}R = \frac{16\pi G}{c^2} ( \rho + \rho_\Lambda )
\end{equation}
where $^{(3)}R$ is the intrinsic curvature, $\rho$ is the matter density and $\rho_\Lambda$ is the density corresponding to a (positive or negative) cosmological constant $\Lambda$ given by equation (\ref{eq:lambda}).
In terms of the radial-radial component of the metric $g_{rr}$ we have
\begin{equation}\label{eq:intcur}
	^{(3)}R = \frac{2}{r^2}\frac{d}{dr}\left\lbrace r(1-g_{rr}^{-1})\right\rbrace
\end{equation}
Using equation \eref{eq:intcur} to integrate \eref{eq:ivalue}, we get:
\begin{equation}\label{eq:grr}
	g_{rr} = \left( 1 - \frac{2Gm(r)}{rc^2} - \frac{8\pi G}{3 c^2}\rho_\Lambda r^2\right)^{-1}
\end{equation}
with
\begin{equation}\label{eq:mr}
	m(r) = \int_0^r \rho(\tilde{r})4\pi \tilde{r}^2d\tilde{r} 
\end{equation}
\end{itemize}
\indent These are our assumptions. Let $R$ denote the radius at which the pressure vanishes \cite{ov}, so that it defines the edge of the fluid sphere. Using spherical symmetry and equation \eref{eq:grr} in the proper volume element, the entropy can be written as:
\begin{equation}\label{eq:entropy}
	S = \int_0^R s(r)\left( 1 - \frac{2Gm(r)}{rc^2} - \frac{8\pi G}{3 c^2}\rho_\Lambda r^2\right)^{-\frac{1}{2}} 4\pi r^2 dr
\end{equation}
Similarly, the baryon number is given by
\begin{equation}\label{eq:number}
	N = \int_0^R n(r)\left( 1 - \frac{2Gm(r)}{rc^2} - \frac{8\pi G}{3 c^2}\rho_\Lambda r^2\right)^{-\frac{1}{2}} 4\pi r^2 dr
\end{equation}
In contrast, to obtain the total mass-energy $M$, including that of the gravitational field, one should {\it not} integrate on the proper volume but on the normal volume (see p. 302 in Weinberg \cite{weinberg}) as in equation \eref{eq:mr}
\begin{equation}\label{eq:mass}
	M \equiv m(R) = \int_0^R \rho(r) 4\pi r^2 dr
\end{equation}
\indent We want to find the extrema of entropy under the constraints of fixed energy and baryon number, so that we work in the microcanonical ensemble. We will use the method of Lagrange multipliers. Let $\beta_0$, $\alpha$ be two, yet undetermined, Lagrange multipliers. The requirement of entropy extrema under the constraints of constant energy and baryon number is given by
\begin{eqnarray}\label{eq:ds1}
	\delta S - \beta_0 c^2 \delta M + \alpha \delta N = 0 \Leftrightarrow \nonumber \\
\fl	\int_0^R\delta s {(g_{rr})}^{\frac{1}{2}} r^2dr + \frac{G}{c^2}\int_0^R s\delta m {(g_{rr})}^{\frac{3}{2}} r dr 
	- \beta_0 c^2 \int_0^R \delta\rho r^2 dr \nonumber\\
	+ \int_0^R \alpha\delta n {(g_{rr})}^{\frac{1}{2}} r^2 dr 
	+ \frac{G}{c^2}\int_0^R \alpha n\delta m {(g_{rr})}^{\frac{3}{2}} r dr	 = 0
\end{eqnarray}
Using
\begin{equation}\label{eq:ds}
	\delta s = \frac{c^2}{T}\delta\rho - \frac{\mu}{T} \delta n
\end{equation}
in the first term and Gibbs-Duhem relation \eref{eq:gibbs} to substitute $s$ in the second term, equation \eref{eq:ds1} becomes
\begin{eqnarray}\label{eq:ds2}
\fl	c^2\int_0^R \frac{1}{T}\delta \rho {(g_{rr})}^{\frac{1}{2}} r^2dr 
		+ \frac{G}{c^2}\int_0^R \frac{\rho c^2 + p}{T}\delta m {(g_{rr})}^{\frac{3}{2}} r dr 
		- \beta_0 c^2 \int_0^R \delta\rho r^2 dr \nonumber\\
\fl		+ \int_0^R \left(\alpha-\frac{\mu}{T}\right) \delta n {(g_{rr})}^{\frac{1}{2}} r^2dr
		+  \frac{G}{c^2}\int_0^R \left(\alpha-\frac{\mu}{T}\right) n \delta m {(g_{rr})}^{\frac{3}{2}} r dr
		= 0
\end{eqnarray}
Since $\rho$ and $n$ are independent variables and $\delta m \sim \delta\rho$, the fourth term of the last equation \eref{eq:ds2} implies
\begin{equation}\label{eq:lmult}
	\alpha = \frac{\mu}{T} 
\end{equation}
This is an expected result, since the temperature and chemical potential should suffer the same red-shift. Using this value for $\alpha$ and equation \eref{eq:mr} we get
\begin{eqnarray}\label{eq:ds3}
\fl	c^2\int_0^R \frac{1}{T}\delta \rho {(g_{rr})}^{\frac{1}{2}} r^2dr 
		+ \frac{G}{c^2}\int_0^R \frac{\rho c^2 + p}{T}{(g_{rr})}^{\frac{3}{2}} r \left(\int_0^r \delta\rho(\tilde{r})4\pi\tilde{r}^2d\tilde{r}\right)  dr \nonumber\\
		- \beta_0 c^2 \int_0^R \delta\rho r^2 dr = 0 \Leftrightarrow \nonumber \\
\fl	\int_0^R dr(\delta\rho) (c^2 r^2)\cdot\left\lbrace
	\frac{1}{T} {(g_{rr})}^{\frac{1}{2}} 
	+ \frac{4\pi G}{c^4}\int_r^R \frac{\rho(\tilde{r}) c^2 + p(\tilde{r})}{T(\tilde{r})}{(g_{rr}(\tilde{r}))}^{\frac{3}{2}} \tilde{r} d\tilde{r}
	- \beta_0	
\right\rbrace
	= 0
\end{eqnarray}
where, going from the first line to the second, we have interchanged in the double integral the integration variables, changing the limits of integration accordingly. In order for this equation to hold for any variation $\delta \rho$, it should be
\begin{equation}\label{eq:b0}
	\beta_0 = \frac{1}{T} {(g_{rr})}^{\frac{1}{2}} 
	+ \frac{4\pi G}{c^4}\int_r^R \frac{\rho(\tilde{r}) c^2 + p(\tilde{r})}{T(\tilde{r})}{(g_{rr}(\tilde{r}))}^{\frac{3}{2}} \tilde{r} d\tilde{r}
\end{equation}
which for $T_0 = 1/\beta_0$ and $r=R$ becomes
\begin{equation}\label{eq:tolman}
\fl	T_0 = T(R){(g_{rr}(R))}^{-\frac{1}{2}} \quad \mbox{or} \quad 
	T_0 = T(R) \left( 1 - \frac{2GM}{Rc^2} - \frac{8\pi G}{3 c^2}\rho_\Lambda R^2\right)^{\frac{1}{2}}
\end{equation}
which is the well known Tolman relation \cite{tolman}. Therefore the Lagrange multiplier $T_0 = 1/\beta_0$ is the surface temperature at the edge $r=R$, measured by an observer at infinity. \\
\indent Let us repeat the proof of a relation given in Gao's Addendum \cite{gao}, that we will use here, as well. By differentiating the Gibbs-Duhem relation \eref{eq:gibbs} and using the first law \eref{eq:1stlaw}, we get
\begin{equation}\label{eq:diffp}
	dp = sdT + nd\mu \Rightarrow p' = s T' + n \mu'
\end{equation}
Equation \eref{eq:lmult} gives
\begin{equation}\label{eq:lagr}
	\mu' = \alpha T'
\end{equation}
Combining equations \eref{eq:diffp}, \eref{eq:lagr} and \eref{eq:gibbs} we have
\begin{equation}\label{eq:Tprime}
	\frac{T'}{T} = \frac{p'}{p+\rho c^2}
\end{equation}
This is the desired equation. Now, differentiating equation \eref{eq:b0}, we get
\begin{eqnarray}
\fl -\frac{T'}{T^2}\left( 1 - \frac{2Gm(r)}{rc^2} - \frac{8\pi G}{3 c^2}\rho_\Lambda r^2\right)^{-\frac{1}{2}}
	-\frac{1}{T}\left( 1 - \frac{2Gm(r)}{rc^2} - \frac{8\pi G}{3 c^2}\rho_\Lambda r^2\right)^{-\frac{3}{2}} \nonumber\\
	\times \frac{1}{c^2}\left(-4\pi G\rho r + \frac{Gm}{r^2} - \frac{8\pi G}{3}\rho_\Lambda r\right) \nonumber\\
	-\frac{1}{T} \frac{1}{c^2}(4\pi G \frac{p}{c^2}r + 4\pi G\rho r)\left( 1 - \frac{2Gm(r)}{rc^2} - \frac{8\pi G}{3 c^2}\rho_\Lambda r^2\right)^{-\frac{3}{2}} = 0
\end{eqnarray}
which after replacing equation \eref{eq:Tprime} gives finally the TOV equation
\begin{equation}\label{eq:TOV2}
\fl	p' = -(\frac{p}{c^2} + \rho)\left( \frac{Gm(r)}{r^2} + 4\pi G \frac{p}{c^2}r - \frac{8\pi G}{3}\rho_\Lambda r\right)
	\left( 1 - \frac{2Gm(r)}{rc^2} - \frac{8\pi G}{3 c^2}\rho_\Lambda r^2\right)^{-1}
\end{equation}

\indent Let us discuss the case of the canonical ensemble. Imagine the fluid sphere to be bounded by reflecting and insulating walls and let on the outside be a heat reservoir with constant temperature $T_0$ at the outer surface of the walls, as measured by an observer at infinity. Since the exchange of heat energy between the system and the reservoir happens only at the edge, the temperature that will enter the expression of the Helmholtz free energy $F$ is $T_0$:
\begin{equation}\label{eq:free}
	F = Mc^2 - T_0 S
\end{equation}
The minima of $F$ will give the stable equilibria. Sometimes is useful (and it is equivalent) to use the Massieu function $J = -\beta_0 F$, i.e.
\begin{equation}\label{eq:massieu}
	J = S - \beta_0 Mc^2 
\end{equation}
Now, the maxima of $J$ will give the stable equilibria. Hereafter, we will refer to the expression \eref{eq:massieu} as the free energy. Since $\beta_0$ is constant and is the inverse temperature at the edge of the sphere as measured by an observer at infinity, it is trivial to see that the extremum, i.e. the point of elimination of the first variation, of the free energy with constant baryon number is equivalent to the extremum of entropy with constant energy and baryon number
\begin{equation}
	\delta J + \alpha \delta N = 0 \Leftrightarrow \delta S - \beta_0 c^2 \delta M + \alpha \delta N = 0
\end{equation}
Thus, the two ensembles, canonical and microcanonical, give the same equilibria, i.e. TOV equation, as they should do. But, the second variation of $J$ and $S$ is different, so that the stability properties of the two ensembles are different, as we will see in the following section.

\section{Thermodynamical stability}\label{sec:thermo}

\indent It is well known that in gravity the thermodynamic ensembles are not equivalent \cite{padman,katz}. Although they describe the same equilibria, the stability properties are different. In general, self-gravitating systems are more unstable in the canonical ensemble rather than the microcanonical. This means that in the microcanonical ensemble there are more equilibrium configurations rather than in the canonical. The question raised is how dynamical stability is related to the thermodynamical stability. In the case of Newtonian gravity, for a spherically symmetric system, it has been proved by Chavanis \cite{chavanisCanonical} that the Jeans dynamical instability is equivalent with the thermodynamical instability in the \textit{canonical} ensemble and very recently a modified Jeans dynamical instability, where the energy is held fixed during the perturbation, is found to be equivalent to the microcanonical thermodynamical instability \cite{Sormani}. In General Relativity, SWZ \cite{swz} have proven for the specific case of radiation that dynamical stability coincides with thermodynamical stability in the \textit{microcanonical} ensemble. The question raised and addressed in this section is whether dynamical stability coincides with microcanonical thermodynamical stability for any equation of state of a perfect fluid in General Relativity. \\
\indent In the microcanonical ensemble, the stability of a equilibrium is determined by the second variation of entropy. If it is negative, it corresponds to a local entropy maxima and the equilibrium is stable, or better metastable in our case since the maxima will be only local entropy maxima. Of course, the global entropy maxima is given by the Bekenstein-Hawking bound at semi-classical level, while at classical level there is no global entropy maxima. Let us calculate the second variation of entropy. \\
\indent Recall that (equations \eref{eq:f}, \eref{eq:drho}, \eref{eq:grr})
\begin{equation}\label{eq:recall}
	f = \delta m\; ,\; \delta\rho = \frac{1}{4\pi r^2}\frac{df}{dr}\; , \; 
	g_{rr} = \left( 1 - \frac{2Gm(r)}{rc^2} - \frac{8\pi G}{3 c^2}\rho_\Lambda r^2\right)^{-1}
\end{equation}
and $\delta m(0) =\delta m(R) = 0$. Also, equation \eref{eq:Tprime} implies
\begin{equation}
	\frac{\delta T}{T} = \frac{\delta p}{p + \rho c^2}
\end{equation}
so that
\begin{equation}\label{eq:deltaT}
	\delta\left(\frac{1}{T}\right) = -\frac{1}{T}\frac{\delta p}{p+\rho c^2}
\end{equation}
Using equations \eref{eq:recall}, \eref{eq:deltaT}, the second variation of entropy given in \eref{eq:entropy} is:
\begin{equation}
\begin{array}{lll}\label{eq:d2S_1}
	\delta^2S = & -\int_0^R dr\frac{c^2}{T}\frac{\delta p}{p+\rho c^2}\frac{df}{dr}{(g_{rr})}^{\frac{1}{2}} &(I_1) \\
	& & \\
	&+\int_0^R dr\frac{2G}{T}\frac{1}{r} f\frac{df}{dr} {(g_{rr})}^{\frac{3}{2}} & (I_2) \\
	& & \\
	&+\int_0^R dr\frac{3G}{T}(p+\rho c^2)\frac{4\pi G}{c^4} f^2 {(g_{rr})}^{\frac{5}{2}} & (I_3)
\end{array}
\end{equation}
where all quantities (except from variations) correspond to an equilibrium, so that the suffix $e$ used in section \ref{sec:dynamical} is suppressed. Integrating by parts integral $(I_2)$ we get
\[
\fl	I_2 = \int_0^R dr \frac{G}{Tr}{(g_{rr})}^{\frac{3}{2}}\left\lbrace \frac{p'}{p+\rho c^2} + \frac{1}{r} + 3\frac{1}{c^2}\left(  -4\pi G \rho r + \frac{Gm}{r^2} - \frac{8\pi G}{3} \rho_\Lambda r\right){g_{rr}} \right\rbrace f^2
\]
Therefore
\[
\fl I_2 + I_3 =  \int_0^R dr \frac{G}{Tr}{(g_{rr})}^{\frac{3}{2}}
		\left\lbrace \frac{p'}{p+\rho c^2} + \frac{1}{r} + 3\frac{1}{c^2}\left(  4\pi G \frac{p}{c^2} r + \frac{Gm}{r^2} - \frac{8\pi G}{3} \rho_\Lambda r\right){g_{rr}} \right\rbrace f^2
\]
which, by use of TOV equation \eref{eq:TOV1} becomes:
\begin{equation}\label{eq:I2I3}
	I_2 + I_3 =  \int_0^R dr \frac{G}{Tr}{(g_{rr})}^{\frac{3}{2}}
		\left( -\frac{2p'}{p+\rho c^2} + \frac{1}{r} \right) f^2
\end{equation}
Integrating by parts the integral $(I_1)$ we get
\begin{eqnarray}
\fl I_1 =  \int_0^R dr \frac{c^2}{T}\frac{1}{p+\rho c^2}{(g_{rr})}^{\frac{1}{2}}
		\left\lbrace {\delta p}' - \delta p \frac{2p' + \rho 'c^2}{p+\rho c^2}\right\rbrace f \nonumber \\
		 - \int_0^R dr \frac{c^2}{T}\frac{1}{p+\rho c^2}{(g_{rr})}^{\frac{3}{2}}
		 \delta p\frac{1}{c^2}\left(  -4\pi G \rho r + \frac{Gm}{r^2} - \frac{8\pi G}{3} \rho_\Lambda r\right) f\nonumber
\end{eqnarray}
which, by use of TOV equation \eref{eq:TOV1} in the second line, becomes:
\begin{eqnarray}\label{eq:I1}
\fl I_1 =  \int_0^R dr \frac{c^2}{T}\frac{1}{p+\rho c^2}{(g_{rr})}^{\frac{1}{2}}
		\left\lbrace {\delta p}' - \delta p \frac{p' + \rho 'c^2}{p+\rho c^2}
		+ \delta p\frac{4\pi G}{c^4}(p+\rho c^2) g_{rr} \right\rbrace f
\end{eqnarray}
Substituting equations \eref{eq:I2I3}, \eref{eq:I1} into \eref{eq:d2S_1} we get
\begin{eqnarray}
	\delta^2 S = \int_0^R dr \frac{1}{T} \frac{c^2}{p+\rho c^2} {(g_{rr})}^{\frac{1}{2}} f \nonumber \\
\fl\label{eq:d2S_2}		
	\times\left\lbrace {\delta p}' - \delta p \frac{p' + \rho 'c^2}{p+\rho c^2}
		+ (p+\rho c^2) g_{rr} \left[\delta p\frac{4\pi G}{c^4} + \frac{G}{rc^2}\left( -\frac{2p'}{p+\rho c^2} + \frac{1}{r}\right)f\right] \right\rbrace 
\end{eqnarray}
Comparing this with equation \eref{eq:instgen} we see that for the quantity in brackets in the above equation to be equal to the left-hand side of \eref{eq:instgen}, and thus the microcanonical thermodynamical stability to be equivalent to linear dynamical stability it must hold:
\begin{equation}
	p'\delta \rho = \delta p \rho '
\end{equation}
For a general equation of state $p = p(\rho)$ the above equation holds. Setting $g = \partial p/\partial \rho c^2$ it is straightforward to check that equation \eref{eq:d2S_2} becomes:
\begin{eqnarray}\label{eq:d2S_3}	
	\delta^2 S = \int_0^R dr \frac{1}{T 4\pi r^2} \frac{c^2}{p+\rho c^2} {(g_{rr})}^{\frac{1}{2}} f \nonumber \\
\fl	\times\left\lbrace gc^2\left[ \frac{d^2}{dr^2} + \left( \frac{g'}{g} - \frac{2}{r}
	 - \frac{(g+1)c^2{\rho}'}{p+\rho c^2} 
		+ \frac{4\pi G}{c^4}r g_{rr}
		(p+\rho c^2)\right)\frac{d}{dr}\right]\right. \nonumber \\
		 + \left. \frac{4\pi G }{c^2}(p+\rho c^2)r g_{rr}\left( -\frac{2gc^2\rho'}{p+\rho c^2} + \frac{1}{r}\right) \right\rbrace f
\end{eqnarray}
The sign of $\delta^2 S$ is therefore determined by the sign of the eigenvalues of the operator
\begin{eqnarray}\label{eq:theoper} 
\fl	\hat{L} = gc^2\left[ \frac{d^2}{dr^2} + \left( \frac{g'}{g} - \frac{2}{r}
	 - \frac{(g+1)c^2{\rho}'}{p+\rho c^2} 
		+ \frac{4\pi G}{c^4}r g_{rr}
		(p+\rho c^2)\right)\frac{d}{dr}\right] \nonumber \\
		 + \frac{4\pi G }{c^2}(p+\rho c^2)r g_{rr}\left( -\frac{2gc^2\rho'}{p+\rho c^2} + \frac{1}{r}\right)
\end{eqnarray}
This operator is exactly equal to the one in equation \eref{eq:dynoper}. The equilibrium is stable if $\hat{L}$ has only negative eigenvalues since then it corresponds to an entropy maxima. The same was true for $\hat{L}$ in the dynamical case in section \ref{sec:dynamical}. Thus, the condition for microcanonical thermodynamical stability is equivalent to the condition for linear dynamical stability. \\
\indent Let us investigate the canonical ensemble as well. For $p = p(\rho)$, the second variation of the free energy \eref{eq:massieu} is:
\begin{eqnarray}\label{eq:d2J}
	\delta^2 J &=& \int_0^R dr\frac{G}{Tr}(g+2)f\frac{df}{dr}{(g_{rr})}^{\frac{3}{2}} \nonumber \\
	&& + \int_0^R dr \frac{3G}{T}(p+\rho c^2) \frac{4\pi G}{c^4} f^2 {(g_{rr})}^{\frac{5}{2}}
\end{eqnarray}
Calculating the first integral by integrating by parts and substituting it in the above equation we get
\begin{eqnarray}\label{eq:d2J}
\fl	\delta^2 J = \int_0^R dr\frac{G}{Tr} {(g_{rr})}^{\frac{3}{2}} f^2\left\lbrace
	-\frac{g'}{2} + \frac{g+2}{2r} + \frac{3g}{2c^2}g_{rr}\left( \frac{Gm}{r^2} - \frac{8\pi G}{3}\rho_\Lambda r\right) 
		\right.\nonumber \\
	\left. + \frac{3}{c^2}g_{rr} \left( 4\pi G\frac{p}{c^2}r + \frac{Gm}{r^2} - \frac{8\pi G}{3}\rho_\Lambda r 
		- 4\pi G\rho r \right)\right\rbrace
\end{eqnarray}
Since $g_{rr} > 0$ (so that no black hole is formed) we see that for a linear equation of state
\[
	p = q\rho c^2 \; , \; q = const.
\]
and for $\Lambda \leq 0$ it is $\delta^2 J > 0$ for any perturbation $f$. Therefore, in this case, the canonical ensemble is completely unstable. 

\section{Newtonian limit}\label{sec:newtonian}

\indent The Newtonian limit of TOV equation is Emden equation and it is taken assuming dust particles \cite{Chandra,agr}. In this case, the equation of state is linear with
\begin{equation}\label{eq:stateN}
	p = \frac{kT}{m_p}\rho 
\end{equation}
where $m_p$ is the mass of one particle. Defining 
\begin{equation}\label{eq:qNewt}
	q = \frac{kT}{m_p c^2}
\end{equation} 
the equation of state can be written as
\begin{equation}\label{eq:stateq}
	p = q\rho c^2 
\end{equation}
and the non-relativistic dust particles limit corresponds to $q \rightarrow 0$. \\
\indent Let us calculate the operator $\hat{L}$ (\ref{eq:theoper}) in the Newtonian limit. Using
$
	g_{rr} \overset{q\rightarrow 0}{\longrightarrow}  1 
$,
equations (\ref{eq:qNewt}), (\ref{eq:stateq}) in order to have only $q$ and not $c$ in $\hat{L}$, and then taking the limit $q\rightarrow 0$, we have:
\begin{eqnarray}
\fl	\hat{L} = \frac{kT}{m_p}\left[ \frac{d^2}{dr^2} + \left( - \frac{2}{r}
	 - \frac{{\rho}'}{\rho} 
		+ q\frac{4\pi G m_p}{kT}r \rho (q+1)\right)\frac{d}{dr}\right] \nonumber \\
		 + \, 4\pi G (q+1)r \rho\left( -\frac{2q\rho'}{(q+1)\rho} + \frac{1}{r}\right) \Rightarrow\nonumber \\
	\hat{L} \overset{q\rightarrow 0}{\longrightarrow} 4\pi\rho r^2 T\left\lbrace  \frac{k}{m_p} \frac{d}{dr}\left( \frac{1}{4\pi\rho r^2}\frac{d}{dr}\right)
	+ \frac{G}{Tr^2} \right\rbrace 
\end{eqnarray}
Therefore, in the Newtonian limit, stability is determined by the eigenvalues of the operator
\begin{equation}\label{eq:newoper}
	\hat{L}_N =  \frac{k}{m_p} \frac{d}{dr}\left( \frac{1}{4\pi\rho r^2}\frac{d}{dr}\right)	+ \frac{G}{Tr^2}
\end{equation}
that is exactly the operator that defines \textit{canonical} thermodynamical stability in the Newtonian Gravity \cite{chavanisCanonical, agrN}. We reach the rather strange and intriguing result that the \textit{microcanonical} ensemble in General Relativity transforms to the \textit{canonical} ensemble in the Newtonian limit! We speculate this is a hint for the presence of an intrinsic heat bath in General Relativity.

\section{Gravothermal catastrophe in General Relativity with $\Lambda$}\label{sec:lambda}

\begin{figure}[tb]
\begin{center}
	\subfigure[$q = 1/3$]{ \label{fig:NvsY_R}\includegraphics[scale=0.42]{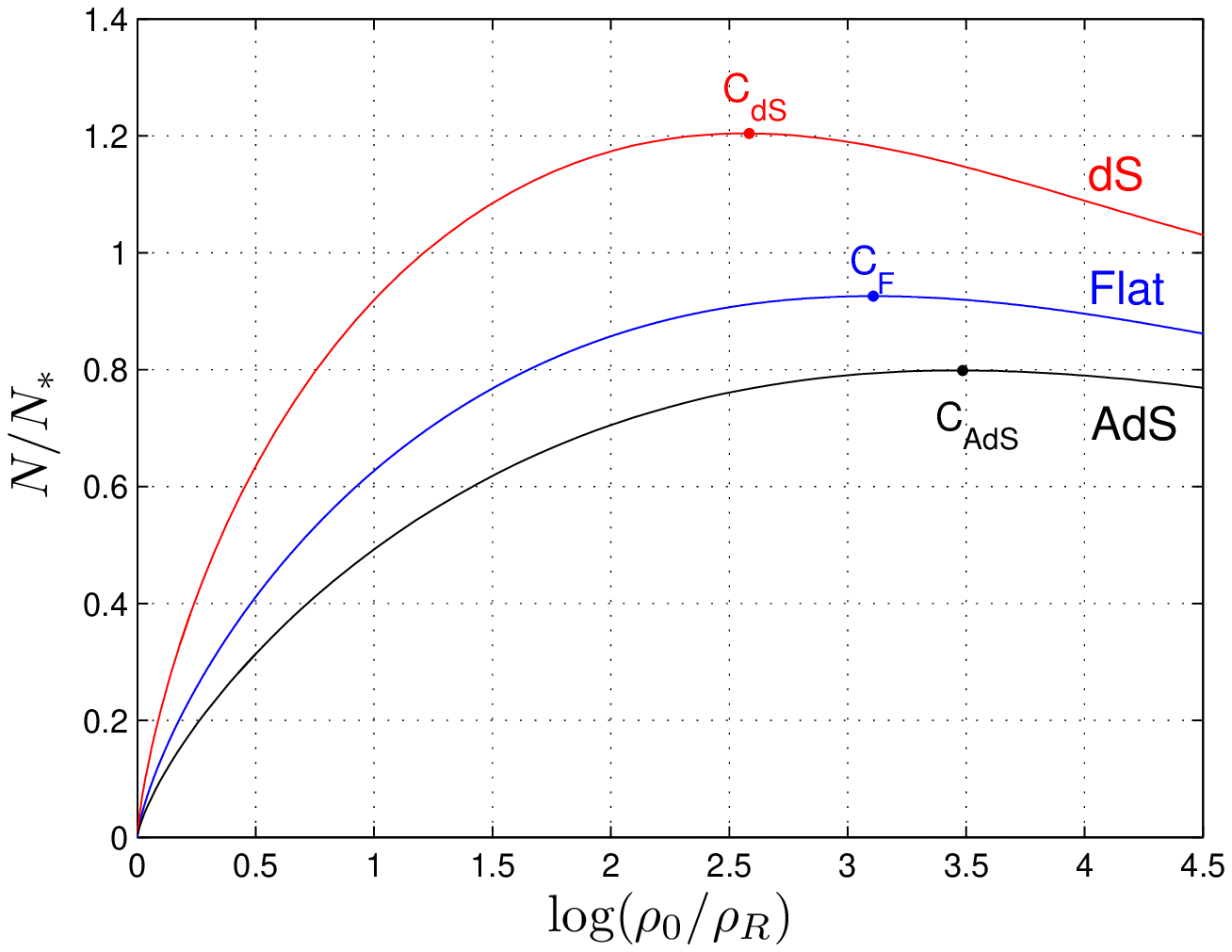} } 
	\subfigure[$q = 1$]{ \label{fig:NvsY_S}\includegraphics[scale=0.42]{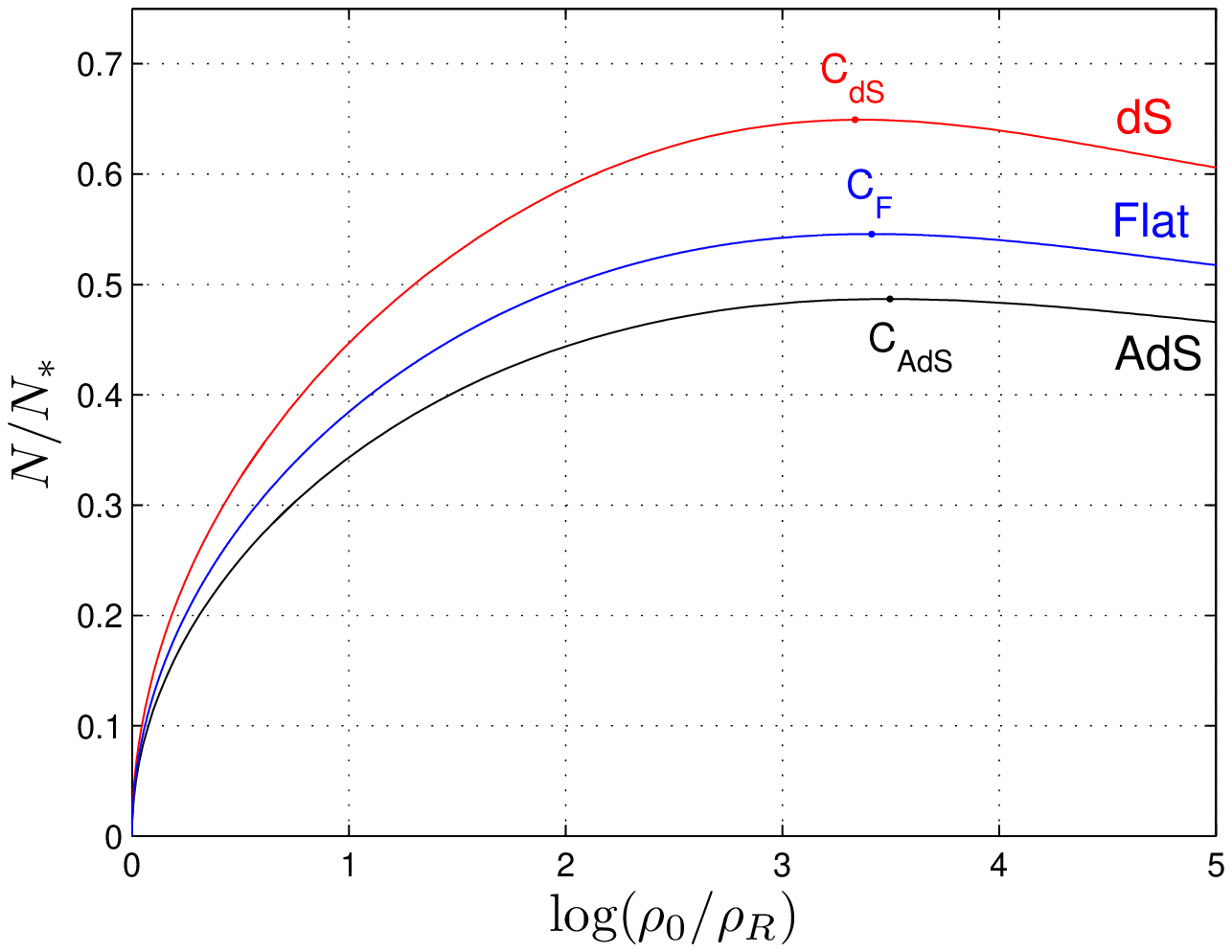} }
\caption{ The baryon number $N$ with respect to the density contrast $\log(\rho_0/\rho_R)$ for radiation ($q=1/3$) and for stiff matter ($q=1$) for asymptotically flat, de Sitter and anti-de Sitter cases. The points $C_i$ are turning points of stability, i.e. at their left side the equilibria are stable, while at their right side, unstable. At each case above $C_i$ there are no equilibria at all.
\label{fig:NvsY}}
\end{center}
\end{figure}

\indent We are interested in this section on the effect of the cosmological constant to the thermodynamical stability of self-gravitating gas. Our interest is mainly theoretical and we basically want to see qualitatively and quantitatively how various -positive or negative- values of the cosmological constant affect the thermodynamical stability. However, for the case of negative cosmological constant the main reason for studying this problem is AdS/CFT and for the case of a positive cosmological constant, apart from theoretical there is also physical justification for studying this problem. Although, at the present epoch of the evolution of the Universe the cosmological constant is considered to be very small, in cosmological models with a time-varying cosmological constant (decaying vacuum) \cite{waga, woodard, polyakov} it is assumed to have been much bigger in the past so that it could have affected the formation of stars during the evolution of the Universe. In addition, in some types of stars, e.g. boson stars, there is an effective cosmological constant, i.e. a term in Einstein's equations similar to the cosmological constant \cite{compact1, compact2, compact3} with a value, big enough to significantly affect the star's configuration. Therefore, the study of stability of TOV for various values of the cosmological constant is of immediate interest for Astrophysics. \\
	\indent This work can be considered as the completion of previous works in the Newtonian case \cite{agr,agrN,agrP} (see also Ref. \cite{AxFlor} for some dynamical effects of the cosmological constant). Let us recall that gravothermal catastrophe \cite{agr,Antonov,Bell-Wood} is called the set of thermodynamic instabilities in the microcanonical ensemble of a, bounded in a spherical box, self-gravitating gas in Newtonian gravity. Recall that the Newtonian limit of TOV is Emden equation \cite{Chandra,agr}. It can also be derived by entropy extremization, so that it describes thermodynamic equilibria as well, just like TOV equation (see section \ref{sec:first}). As we have seen for a linear equation of state
\begin{equation}\label{eq:state}
	p = q\rho c^2 \; , \; q = const.
\end{equation}
the dust particles limit corresponds to $q \rightarrow 0$. In the present study we assume this linear equation of state for $0<q\leq 1$. We will also assume a constant chemical composition and a constant entropy per nucleon ($d(s/n) = 0$) throughout the entire fluid sphere. These assumptions correspond to the following physical cases \cite{weinberg}: (a) a white dwarf or a neutron star of low mass, in which the temperature is essentially at absolute zero (Nernst's theorem then ensures $s/n = 0$); (b) stars in convective equilibrium, like supermassive stars and self-gravitating radiation (`photon stars'). Another interesting case is the `stiff' case of matter for which $q = 1$ where the speed of sound equals the speed of light (e.g. black hole gas in Banks-Fischler cosmology \cite{banks-fischler}). \\
\indent The problem of dynamical stability under the same assumptions in the asymptotically flat case has been studied by Chavanis \cite{chavanisR1,chavanisR2}. He uses Weinberg's theorems (p. 305 in \cite{weinberg}) which state that, under the above assumptions, in a series of equilibria, the mass as long as the baryon number have a maximum at a turning point of stability. He proves that non-linear dynamical stability as defined in Weinberg coincides with linear dynamical stability for matter with a linear equation of state and with constant entropy per nucleon. SWZ \cite{swz} had proven for the specific case of radiation that linear dynamical stability coincides with microcanonical thermodynamical stability. We have proven in the previous section that, in the most general case under no specific assumptions, linear dynamical stability coincides with microcanonical thermodynamical stability. So that, all cases studied by Chavanis in \cite{chavanisR2} are in fact gravothermal instabilities. In addition this equivalence we have proven, hints to a relation between the stability theorems of Weinberg with the one of Poincar\'e \cite{Poincare, Katz1}. In fact it seems that Weinberg's theorem is just a specific case of Poincar\'e's theorem of linear series of equilibria. \\
\indent Here, we are mainly interested in the type of instability that is related with the complete absent of equilibria. That is, with the region above the mass or baryon number maximum in a series of equilibria. This region is characterized by a minimum radius down to which equilibrium configurations do exist. The other type of instability is a `weaker' one and is related with equilibrium configurations, i.e. true equilibria in a series of equilibria, which nevertheless are unstable, i.e. the entropy is a saddle point for these equilibria. However, this instability is also characterized by the same point, i.e. the maximum of $M$ and $N$ (which coincide as Weinberg has proven) at which the stable equilibria become unstable. This point is called `a turning point of stability'. So that, practically our scope is to determine the maximum of $N$ for various values of $\Lambda$ in a series of equilibria, i.e. equilibria for various density contrasts $\log(\rho_0/\rho_R)$ where $\rho_0$ is the density at the center and $\rho_R$ at the edge of the sphere. \\
\indent Let us prove a relation that we will need in the followings. The first law of thermodynamics can be expressed in the form:
\begin{equation}
	Td(\frac{s}{n}) = d(\frac{\rho c^2}{n}) + pd(\frac{1}{n})
\end{equation}
Under the above assumptions the term $Td(s/n)$ can be neglected, so that we get:
\begin{equation}
	d\rho c^2 = \frac{p+ \rho c^2}{n}d n
\end{equation} 
Integrating the above equation, taking into account the equation of state \eref{eq:state}, we get the polytropic equation:
\begin{equation}\label{eq:polyt}
	p = K n^{q+1} 
\end{equation} 
Combining the polytropic equation \eref{eq:polyt} with the equation of state \eref{eq:state} we get: 
\begin{equation}
	n = \left(\frac{q}{K} \rho c^2\right)^{\frac{1}{q+1}}
\end{equation}
\indent Now, let our fluid sphere be bounded by perfectly reflecting and non-insulating walls. Let us use the dimensionless variables of Chandrasekhar \cite{Chandra}
\begin{equation}\label{eq:cvar}
	\rho = \rho_0 e^{-y}\; , \;\; x = r \sqrt{4\pi G \rho_0\frac{q+1}{qc^2}}
\end{equation}
where $y$, $x$ correspond to $\psi$, $\xi$ of Chandrasekhar (we preferred to keep the notation relevant to the Newtonian limit \cite{Bell-Wood, agr}) and $\rho_0$ is the mass density at the center of the fluid sphere. Let us introduce also:
\begin{equation}
	\lambda = \frac{2\rho_\Lambda}{\rho_0} \; ,\;\;
	\mu  (x) = \frac{1}{4\pi \rho_0}\left( 4\pi G\rho_0 \frac{q+1}{qc^2} \right)^{\frac{3}{2}}m(r)\; , \;\; M = m(R)
\end{equation}
where $R$ is the radius of the sphere. Using these dimensionless variables, TOV equation \eref{eq:TOV2} becomes:
\begin{eqnarray}
\label{eq:TOV_ND}	\frac{dy}{dx} = \left( \frac{\mu}{x^2} + q x e^{-y} - \frac{\lambda}{3} x\right) 
		\left( 1 - 2\frac{q}{q+1}\frac{\mu}{x} - \frac{\lambda}{3}\frac{q}{q+1} x^2 \right)^{-1} \\
\label{eq:dm_ND}	\frac{d\mu}{dx} = x^2 e^{-y}
\end{eqnarray}
with initial conditions
\begin{equation}
	y(0) = \mu(0) = 0
\end{equation}
We introduce the dimensionless energy:
\begin{equation}\label{eq:Q1}
	Q \equiv \frac{2GM}{Rc^2} = \frac{2\mu}{z}\frac{q}{q+1} 
\end{equation}
Let $z$ denote the value of $x$ at $R$. Integrating TOV \eref{eq:TOV_ND} is straightforward to calculate $\mu$ at $z$ and after substituting it in \eref{eq:Q1} we get:
\begin{equation}\label{eq:Q2}
	Q(z)  = \frac{2q}{q+1}\frac{zy'(z)-qz^2 e^{-y(z)} - \frac{\lambda}{3}z^2\left(\frac{q}{q+1}zy'(z) + 1\right)}{2\frac{q}{q+1}zy'(z) + 1} 
\end{equation}
where $y'(z)$ denotes $\left. \frac{dy}{dx}\right|_{x=z}$. We use the dimensionless baryon number introduced by Chavanis \cite{chavanisR1} which we call $B$:
\begin{equation}\label{eq:BN}
	B = \frac{N}{N_*}\;,\;N_* = 4\pi R^3 \left(\frac{1}{4\pi G K R^2}\frac{q^2c^4}{q+1} \right)^{\frac{1}{q+1}}
\end{equation}
Using equation \eref{eq:number} for $N$ we find in dimensionless variables:
\begin{equation}
	B(z) = \frac{1}{z^{\frac{3q+1}{q+1}}}\int_0^z x^2 e^{-\frac{y}{q+1}} \left(1 - 2\frac{q}{q+1}\frac{\mu}{x}  - \frac{\lambda}{3}\frac{q}{q+1}x^2 \right)^{-\frac{1}{2}}dx 
\end{equation}
For a series of equilibria, where each one is specified by the dimensionless quantity $z$ that determines the radius $R$ of the sphere and/or its central density $\rho_0$, the quantity $B(z)$ expresses the baryon number for a fixed radius, $Q(z)$ expresses the total mass for a fixed radius, while $1/Q(z)$ expresses the radius for a fixed mass. Instead of $z$, it is equivalent to use the so called density contrast $\log(\rho_0/\rho_R)$, as a free variable to define the equilibria.
\\
\indent In the asymptotically flat case it is trivial to generate a series of equilibria, since the system of equations \eref{eq:TOV_ND}, \eref{eq:dm_ND} can be regarded with respect to $z$ instead of $x$ with no problem. However, in the presence of $\Lambda$ it is false to consider equation \eref{eq:TOV_ND} with respect to $z$. There appears the variable $\lambda$ in the equations which cannot be regarded as constant, because it contains $\rho_0$. So that, in fact there are two independent variables $(\lambda,z)$. To overcome this difficulty we developed a strategy similar to the one that has been used in order to solve the corresponding Newtonian problem \cite{agr,agrN}. The cosmological constant introduces a mass scale to the system $M_\Lambda = \rho_\Lambda \frac{4}{3}\pi R^3$. We define the dimensionless mass
\begin{equation}\label{eq:m_ND}
	\tilde{m} \equiv \frac{M}{2M_\Lambda} = \frac{3}{8\pi}\frac{M}{\rho_\Lambda R^3} = \frac{3\mu}{\lambda z^3}
\end{equation}
We solve numerically the system \eref{eq:TOV_ND}-\eref{eq:dm_ND} for a $z$ range of values choosing this value for $\lambda$ that keeps constant the dimensionless mass $\tilde{m}$. This operation is performed by a computer program we developed. In this way a consistent series of equilibria can be generated that corresponds to a fixed $M$ or $R$ or $\rho_\Lambda$ according to equation \eref{eq:m_ND}. Then, by generating series for various $\tilde{m}$ we can see how various critical quantities vary with $\Lambda$.
\\	
\indent In \fref{fig:NvsY} we can see how, the baryon number changes with respect to the density contrast for some fixed values of cosmological constant in a series of equilibria. The fixed positive value (corresponding to $\tilde{m} = 1$) is denoted as `dS' and the fixed negative value (corresponding to $\tilde{m} = -1$) is denoted as `AdS', while the case $\Lambda = 0$ is denoted as `Flat'. The series of equilibria are plotted for the cases of $q=1/3$ (radiation or neutron stars) and $q=1$ (stiff matter). We see that, as the cosmological constant increases, the system is stabilized with respect to the `strong' instability, i.e. the one corresponding to no equilibria, since the turning point goes to higher $N$ values. The `weak' instability, i.e. the one corresponding to unstable equilibria, occurs at lower density contrast values as the cosmological constant increases. \\
\begin{figure}[tb]
\begin{center}
	\subfigure[$q = 1/3$]{ \label{fig:Rcr_R}\includegraphics[scale=0.4]{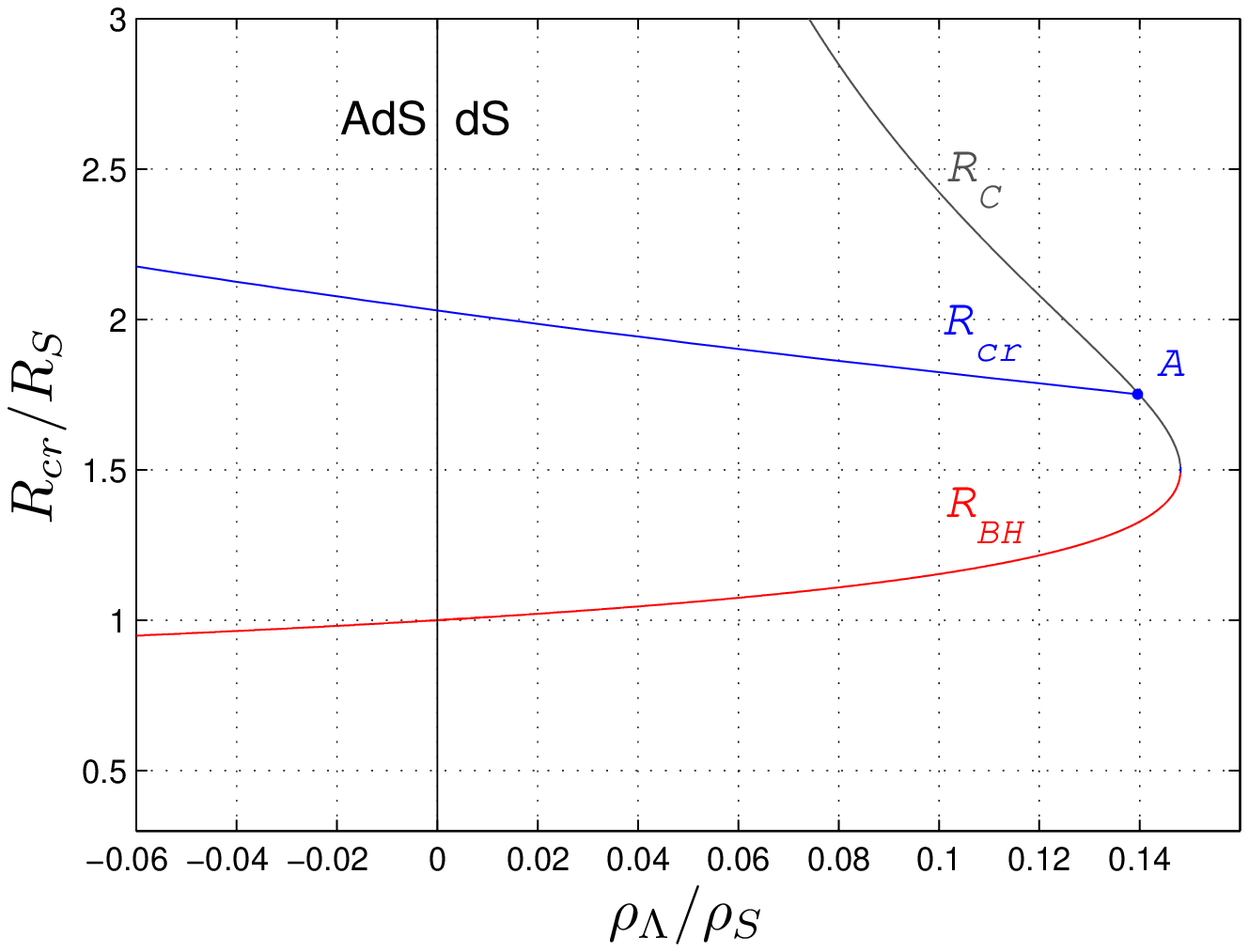} } 
	\subfigure[$q = 1$]{ \label{fig:Rcr_S}\includegraphics[scale=0.4]{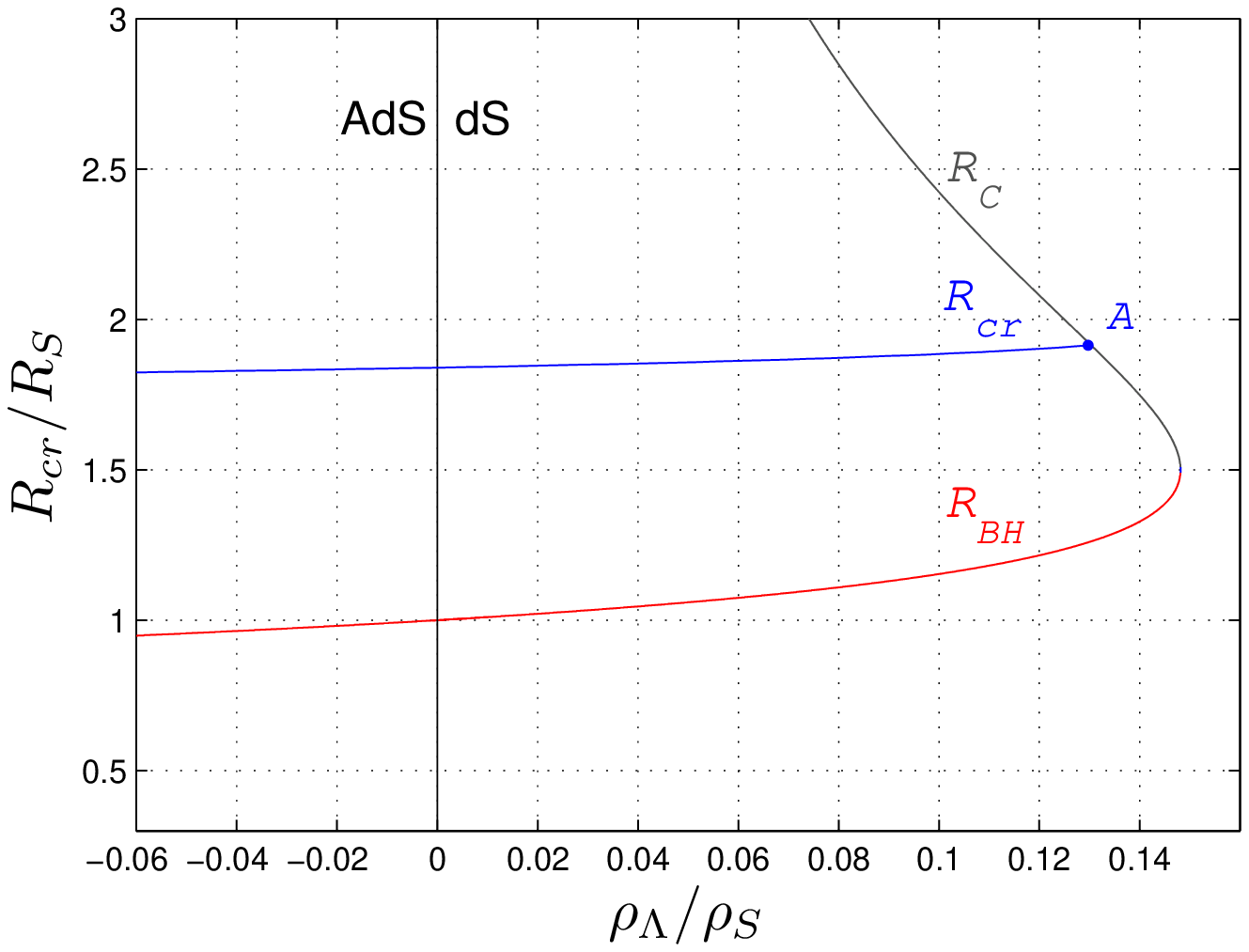} }
\caption{ The critical radius $R_{cr}$ with respect to the cosmological constant $\rho_\Lambda$ in Schwartzschild units $R_S = 2GM/c^2$, $\rho_S = 3M/4\pi R_S^3$ for a fixed mass $M$. The spheres of radius $R < R_{cr}$ are strongly thermodynamically unstable, i.e. there exist no equilibria at all. For $R > R_{cr}$ there exist equilibria which are stable or unstable depending on the density contrast $\rho_0/\rho_R$. In the same plot, are drawn the black hole radius $R_{BH}$ and the cosmological radius $R_C$, as well. At points $A$, the critical radius hits the cosmological horizon.
\label{fig:Rcr}}
\end{center}
\end{figure}
\indent We would like to plot the critical radius, i.e. the minimum radius of the fluid sphere down to which equilibria do exist, with respect to the cosmological constant. However, there is the following complication in this procedure. In an asymptotically de Sitter space, every fluid sphere of mass $M$ and radius $R$ has two characteristic length scales: the black hole radius $R_{BH}$, that would correspond to a black hole horizon if $R < R_{BH}$ and the cosmological radius $R_C$, that would correspond to a cosmological horizon if $R > R_C$ (see \fref{fig:Rcr}). By black hole radius we do not mean the Schwartzschild radius $GM/c^2$ but the larger `Schwartzschild-de Sitter' radius that corresponds to the smaller real root of the polynomial with respect to $R_H$:
\begin{equation}\label{eq:poly}
	1 - \frac{2GM}{R_H c^2} - \frac{8\pi G}{3 c^2}\rho_\Lambda R_H^2 = 0
\end{equation}
The cosmological radius, that we denote $R_C$, is the larger real root of the polynomial \eref{eq:poly}. 
We expect and indeed verify numerically that the critical radius $R_{cr}$ is always larger than $R_{BH}$. Thus, the system becomes unstable always before it reaches its black hole horizon radius and become a black hole. On the contrary, as the cosmological constant is increasing, i.e. $R_C$ is decreasing, one would expect that $R_{cr}$ might become larger than $R_C$. This would mean that the sphere with $R = R_C = R_{cr}$ is the maximum sphere for which an instability can be observed, because for any bigger sphere the critical radius would lie beyond the cosmological horizon. \\
\indent We can determine the dimensionless mass $\tilde{m}$ at which $R = R_C$ with respect to $Q = 2GM/Rc^2$ as follows. It is easy to verify that the polynomial \eref{eq:poly}, for a sphere of radius $R$, and for $R_H = R_C$ can be written in the form:
\begin{equation}\label{eq:poly1}
	1 - Q\frac{1}{(R_C/R)} - \frac{Q}{2\tilde{m}}\left(\frac{R_C}{R}\right)^2 = 0
\end{equation}
We see that $R = R_C$ if:
\begin{equation}\label{eq:conditionC}
	2\tilde{m} = \frac{Q}{1-Q}
\end{equation}
We verify numerically that this condition holds when the critical radius `hits' the cosmological horizon. Physically $R_{cr} > R_C$ is meaningless and this is manifested mathematically (and numerically) as inability to calculate various physical quantities (for example $N$ becomes complex and the computer program breaks down). So that a fluid sphere with $R_{cr} > R_C$ can be considered to be stable, since matter beyond the cosmological horizon cannot interact with matter inside it. In \fref{fig:Rcr} is plotted the critical radius in Schwartzschild units, together with the black hole and cosmological radii, with respect to the cosmological constant for a fixed mass. The critical radius hits the cosmological horizon at point $A$. Spheres with radii smaller than the critical radius are microcanonical thermodynamically unstable, in the sense that there are no thermodynamic equilibria in the microcanonical ensemble. \\
\indent Recall that the Newtonian limit of TOV equation is taken for $q = kT/m_pc^2 \ll 1$. Let us calculate $B = N/N_*$ for dust matter. Substituting $c^2 = 1/qm_p\beta$ into equation \eref{eq:BN}, then taking $q\rightarrow 0$, equation \eref{eq:BN} becomes:
\begin{equation}
	B \overset{q\rightarrow 0}{\longrightarrow} m_p\frac{GM\beta}{R} 	
\end{equation}
This is exactly the Newtonian dimensionless temperature as defined in \cite{agrN,Bell-Wood,chavanisCanonical} for $m_p = 1$. Since $B$ controls microcanonical thermodynamical stability in relativity and canonical thermodynamical stability in Newtonian limit \cite{agrN,chavanisCanonical}, we see once more that the Newtonian limit of the relativistic microcanonical ensemble is the canonical ensemble\footnote{In addition, we verified numerically that for $q=0$, i.e. non-relativistic matter, the critical value of $z$ for any value of the cosmological constant equals the critical $z$ of the Newtonian canonical ensemble (and not of the Newtonian microcanonical ensemble).}, verifying the result of section \ref{sec:newtonian}. This conclusion can also be intuitively justified, since the canonical ensemble in Newtonian Gravity behaves qualitatively similar to a General Relativistic system. That is because in the Newtonian canonical ensemble, for a fixed mass, the thermodynamical instability is triggered as the system is contracted beyond a minimum radius (i.e. for \textit{small} radii) \cite{agrN}, just like in General Relativity, while in the Newtonian microcanonical ensemble, the instability sets in when the system is expanded beyond a maximum radius (i.e. for \textit{big} radii).

\section{Conclusions}

\indent Probably, the most interesting result of this work is the fact that for a static spherically symmetric perfect fluid, the Einstein's equations are completely equivalent with the thermodynamics of self-gravitating gas in the microcanonical ensemble. This equivalence does not apply only on the solutions of Einstein's equations, i.e. the equilibria, but also on the stability of these equilibria! This result gives further support on the deep connection between gravity and thermodynamics. \\
\indent A rather intriguing result we found, is that the \textit{microcanonical} ensemble in General Relativity becomes the \textit{canonical} ensemble in the Newtonian limit (dust matter) as we have analytically calculated in section \ref{sec:newtonian}.
It seems, as though in General Relativity, there is an implicitly constructed `heat bath' so that a microcanonical ensemble would look like a canonical ensemble when non-relativistic matter is considered! \\
\indent Finally, we have studied how the stability properties are influenced by the presence of a cosmological constant. The main results, we could say are summarized in the characteristic figures \ref{fig:NvsY} and \ref{fig:Rcr}. For the cases of radiation and neutron cores, the critical radius less than which there is no thermodynamical equilibria, is decreasing with increasing cosmological constant while for stiff matter it is increasing. This critical radius is always bigger than the black hole radius so that the system becomes unstable long before it reaches its black hole radius for any value of the cosmological constant. However, at some point the critical radius hits the cosmological horizon. Any sphere with bigger radius can be considered stable since matter outside the cosmological horizon cannot interact with matter inside. Moreover, an increase in the cosmological constant tends to stabilize the system thermodynamically, rendering de Sitter space, at least at the purely classical level, more stable than anti-de Sitter. \\
\indent Closing, let us discuss some possible extensions and applications. One would naturally expect that an entropic variational principle could also be applied to rotating relativistic stars in hydrostatic equilibrium (see Refs. \cite{Stergioulas, Gourgoulhon} for recent reviews on rotating relativistic stars). Variational principles of energy have already been developed and are being used in the study of rotating compact stars, starting from the early work of Hartle \& Sharp \cite{Hartle}, while a thermodynamic treatment of rotating bodies and their black hole limit is developed in Ref. \cite{Neugebauer}. We stress out that, nowadays, instabilities of rotating relativistic stars are important and of topical interest in Astrophysics, because of the numerous experiments, that are devoted to detect gravitational waves \cite{Andersson, Villain}. It is reasonable to expect, that a thermodynamic treatment, apart from the conceptual insight that would provide in the study of instabilities of rotating relativistic stars, would also give a technical advantage on determining the stability region of equilibria. This is because in dynamical instabilities specific oscillation modes are studied one by one, while entropic arguments are independent of the specific mode of instability, rendering a thermodynamical instability a broader concept.

\section*{Acknowledgements}

\indent I thank Minos Axenides for his support and guidance and George Georgiou for the plentiful, invaluable discussions that substantially contributed to this work. The work was supported in part by the
``Development Proposals of Research Institutions-KRIPIS", ESPA 2007-2013.

\appendix
\section{}\label{app:A}

\indent We review here Yabushita's \cite{yabushita} calculation of the equation that determines linear stability, including a cosmological constant. Using the metric \eref{eq:metric}, the energy momentum tensor \eref{eq:enmom} has the following components (p. 251 in Tolman \cite{tolman})
\begin{eqnarray}
\label{eq:tensorcom1}	\frac{8\pi G}{c^4}T^0_0 = e^{-\lambda}\left( \frac{\lambda '}{r} - \frac{1}{r^2}\right) + \frac{1}{r^2} - 							\Lambda \\
\label{eq:tensorcom2}	\frac{8\pi G}{c^4}T^1_1 = -e^{-\lambda}\left( \frac{\nu '}{r} + \frac{1}{r^2}\right) + \frac{1}{r^2} - 
							\Lambda \\
\label{eq:tensorcom3}\fl	\frac{8\pi G}{c^4}T^2_2 = \frac{8\pi G}{c^4}T^3_3 = 
		-e^{-\lambda}\left( \frac{\nu ''}{2} - \frac{\lambda'\nu'}{4} + \frac{{\nu '}^2}{4} + \frac{\nu'-\lambda'}{2r}\right) 
		\nonumber \\
		+ e^{-\nu}\left( \frac{\ddot{\lambda}}{2} + \frac{\dot{\lambda}^2}{4} - \frac{\dot{\lambda}\dot{\nu}}{4}\right) 
			- \Lambda \\ 	
\label{eq:tensorcom4}	\frac{8\pi G}{c^4}T^1_0 = -e^{-\lambda} \frac{\dot{\lambda}}{r} \\
\label{eq:tensorcom5}		\frac{8\pi G}{c^4}T^0_1 = e^{-\nu} \frac{\dot{\lambda}}{r} 
\end{eqnarray}
where prime denotes differentiation with respect to $r$ and dot with respect to $t$. The four velocity has components $u^\mu = (u^0,u^1,0,0)$. At the equilibrium $u^1 = 0$, the energy momentum tensor has only the followings non-zero components
\begin{equation}
	T^0_0 = \rho \quad , \quad T^1_1 = T^2_2 = T^3_3 = -p
\end{equation}
and $\lambda$, $\nu$ are independent of $t$. Therefore equations \eref{eq:tensorcom1}-\eref{eq:tensorcom3} become
\begin{eqnarray}
\label{eq:tolequi1}	\frac{dp_e}{dr} = -\frac{1}{2}(p_e + \rho_e c^2) \nu_e' \label{eq:tolequia}\\
\label{eq:tolequi2}	\frac{8\pi G}{c^2}\rho_e = e^{-\lambda_e}\left( \frac{\lambda_e '}{r} - \frac{1}{r^2}\right) + \frac{1}{r^2} 
						- \Lambda \\
\label{eq:tolequi3}	\frac{8\pi G}{c^4} p_e  = e^{-\lambda_e}\left( \frac{\nu_e '}{r} + \frac{1}{r^2}\right) - \frac{1}{r^2} + 
						\Lambda 
\end{eqnarray}
where the suffix $e$ denotes quantities at the equilibrium. Equation \eref{eq:tolequi1} is the relativistic analogue of Newtonian expression for the hydrostatic equilibrium
\begin{equation}
	\frac{dp}{dr} = -\rho\frac{d\phi}{dr}
\end{equation}
It is derived by equating \eref{eq:tensorcom2} and \eref{eq:tensorcom3}. Applying the transformation
\begin{equation}\label{eq:lambda_m}
	e^{-\lambda} = 1 - \frac{2Gm(r)}{rc^2} - \frac{\Lambda}{3}r^2
\end{equation}
to equation \eref{eq:tolequi3}, solving with respect to $\nu_e'$ and substituting $\nu_e'$ into equation \eref{eq:tolequi1}, we get TOV equation:
\begin{equation}\label{eq:TOV1}
\fl	{p_e}' = -(\frac{p_e}{c^2} + \rho_e)\left( \frac{Gm_e(r)}{r^2} + 4\pi G \frac{p_e}{c^2}r - \frac{\Lambda}{3}c^2 r\right)
	\left( 1 - \frac{2Gm_e(r)}{rc^2} - \frac{\Lambda}{3}r^2\right)^{-1}		
\end{equation}
that is the relativistic equation of hydrostatic equilibria. The remaining equation \eref{eq:tolequi2} gives
\begin{equation}\label{eq:mprime}
	m_e' = 4\pi\rho_e r^2
\end{equation}
which means that $m$ is the mass contained inside the radius $r$. Equations \eref{eq:TOV1} and \eref{eq:mprime} together with an equation of state $p_e = p_e(\rho_e)$ fully describe the equilibria, provided some initial conditions are given. \\
\indent Now, let us consider small perturbations about an equilibrium configuration
\begin{equation}\label{eq:pert}
\fl		\lambda = \lambda_e + \delta\lambda \,,\; \nu = \nu_e + \delta\nu \,,\; 
		p = p_e + \delta p \,,\;  \rho = \rho_e + \delta \rho \,,\;
		u^1 = \delta u^1 \,,\; u^0 = u^0_e +\delta u^0
\end{equation}
with all variations depending on both $r$ and $t$. To first order, equations \eref{eq:tensorcom1}-\eref{eq:tensorcom3} yield
\begin{equation}\label{eq:ddot}
	\frac{1}{r}e^{-\nu_e} \ddot{\delta\lambda} = \frac{8\pi G}{c^4}\left\lbrace \delta p' + \frac{\nu'}{2}(\delta p + \delta \rho c^2)
		+\frac{1}{2}(p_e + \rho_e c^2)\delta \nu'\right\rbrace
\end{equation}
The conservation of mass ${T^\mu_0}_{;\mu} = 0$ gives 
\begin{equation}
	\dot{\rho} + {T^1_0}' + \frac{1}{2}\dot{\lambda}(p+ \rho c^2) + T^1_0 \left(\frac{1}{2}\lambda' + \frac{1}{2}\nu' + \frac{2}{r}\right) = 0
\end{equation}
which after replacing \eref{eq:pert} and differentiation with respect to $t$ and use of equations \eref{eq:tolequi1}-\eref{eq:tolequi3} gives
\begin{equation}\label{eq:inter}
	\ddot{\delta\rho} - \frac{c^4}{8\pi G}\left(\frac{\partial}{\partial r} + \frac{2}{r}\right)\frac{e^{-\lambda_e}}{r}\ddot{\delta\lambda} = 0
\end{equation}
Substituting equation \eref{eq:ddot} into \eref{eq:inter} we get
\begin{equation}
\fl	 \ddot{\delta\rho} - \left( \frac{\partial}{\partial r} + \frac{2}{r}\right)
	 \left\lbrace e^{\nu_e-\lambda_e} \left[ \delta p' + \frac{{\nu_e}'}{2}(\delta p + \delta \rho c^2)
		+\frac{{\delta\nu}'}{2}(p_e + \rho_e c^2) \right] \right\rbrace
\end{equation}
Let 
\begin{equation}
	f(r,t) \equiv \delta m(r,t)
\end{equation}
with
\begin{equation}
	\delta \rho = \frac{1}{4\pi r^2}\frac{\partial f}{\partial r}
\end{equation}
Assuming a perturbation
\begin{equation}
	\delta m \sim e^{\sigma t}
\end{equation}
with $\delta m(r=0) = 0$ and integrating equation \eref{eq:ddot} we get finally
\begin{equation}\label{eq:yabu}
 \delta p' + \frac{{\nu_e}'}{2}(\delta p + \delta \rho c^2)
		+\frac{{\delta\nu}'}{2}(p_e + \rho_e c^2)  = \frac{e^{\lambda_e - \nu_e}}{4\pi r^2}\sigma^2 f
\end{equation}
The cosmological constant does not enter explicitly in the equation but only implicitly in the equilibrium quantities.

\section*{References}
% refs.tex                           

\end{document}